\documentclass[aps,pra,twocolumn,groupedaddress,nofootinbib]{revtex4-1}
\usepackage{amsmath, amsfonts, amssymb,amsthm}
\usepackage{graphicx,subfigure}
\usepackage{mathrsfs}
\usepackage[mathscr]{eucal}
\usepackage{times,txfonts}
\usepackage{bm}
\usepackage{subfigure}
\usepackage[bookmarks=false]{hyperref}
\usepackage{xcolor}
\usepackage{makecell}
\hypersetup{colorlinks=true,citecolor=blue,
linkcolor=blue,urlcolor=blue,pdfstartview=FitH,
bookmarksopen=true}

\begin{document}
\title{Basis-independent Coherence in Noninertial Frames}
\author{Ming-Ming Du$^1$}
\email{mingmingdu@njupt.edu.cn}
\author{Yi-Hao Fan$^1$, Hong-Wei Li$^1$, Shu-Ting Shen$^1$, Xiao-Jing Yan$^2$, Xi-Yun Li$^2$,Wei Zhong$^3$,Yu-Bo Sheng$^{1,3}$}
\author{Lan Zhou$^2$}
\email{zhoul@njupt.edu.cn}
\affiliation{$1.$ College of Electronic and Optical Engineering and College of Flexible Electronics (Future Technology), Nanjing
University of Posts and Telecommunications, Nanjing, 210023, China\\
$2.$ School of Science, Nanjing University of Posts and Telecommunications, Nanjing,
210023, China\\
$3.$ Institute of Quantum Information and Technology, Nanjing University of Posts and Telecommunications, Nanjing, 210003, China}
\date{\today}
\begin{abstract}
We investigate the behavior of basis-independent quantum coherence between two modes of a free Dirac field as observed by relatively accelerated observers. Our findings reveal three key results: (i) the basis-independent coherence between modes \(A\) and \(B_{I}\) decreases with increasing acceleration but remains finite even in the limit of infinite acceleration; (ii) at zero acceleration, the coherence between modes \(A\) and \(B_{II}\) is nonzero—contrasting with the behavior of basis-dependent coherence, which typically vanishes in this case; and (iii) the basis-independent coherence between modes \(B_I\) and \(B_{II}\) remains constant regardless of acceleration, exhibiting a freezing phenomenon. These results demonstrate the intrinsic robustness of basis-independent coherence under Unruh effects.
\end{abstract}

\maketitle

\section{Introduction}
Quantum coherence is a fundamental feature of quantum mechanics, characterizing the ability of a quantum state to sustain superposition and give rise to interference phenomena. As one kind of quantum resources \cite{Streltsov2017}, coherence plays a central role in many fields such as quantum computing \cite{Hillery2016,Naseri2022,Ahnefeld2022,Pan2022,Shi2017}, quantum metrology \cite{Giorda2018}, thermodynamics \cite{Lostaglio2015,Lostaglio2015A,Cwiklinski2015,Narasimhachar2015,Korzekwa2016}, and quantum biology \cite{Lloyd2011,Huelga2013}. Out of various physical contexts, researchers have proposed different measures for quantifying coherence \cite{Baumgratz2014,Streltsov2015,Mandel1995,Yu2015,Yu2016A,Zhang2018,Radhakrishnan2019}. One of them is  basis-independent coherence \cite{Radhakrishnan2019}. It is defined as \cite{Radhakrishnan2019}
\begin{align}\label{coherence}
C(\rho) = \sqrt{ S \left( \frac{\rho +\rho_M}{2} \right) - \frac{S(\rho) + \log_2 d}{2}},
\end{align}
where $S(\bullet)$ represents the von Neumann entropy, $\rho_M=\mathrm{I}/d$ is the maximally mixed state in a $d$-dimensional Hilbert space, and $ log_2d$ is the logarithm of $d$ to base $2$. Different from the coherence measures based on the rigorous framework \cite{Baumgratz2014}, which are dependent on the basis of the Hilbert space, the  basis-independent coherence is a basis-independent quantity, describing the intrinsic property of a state. It provides a more universal measure of quantumness for a state \cite{Radhakrishnan2019,Yin2022,Cao2019,Designolle2021}. Physically, basis-independent coherence captures the minimum irreducible coherence present in a quantum state regardless of observer-dependent measurement choices.

The development of measures of quantum coherence has opened up the possibility of exploring the impact of relativistic effects on quantum coherence, which is a crucial step in advancing quantum information processing tasks in extreme physical environments \cite{Liu2025,Liu2025b,Liu2025c,Tang2025,Li2025,Wu20241,Wu20242,Wu20243} and deepening our understanding of black-hole thermodynamics and the black-hole information paradox \cite{Fuentes-Schuller2005,Alsing2006,Bruschi2012,Qiang2015,Bruschi2010,Wang2016,He2018,Huang2018,Kollas2023,Harikrishnan2022,Du2024,Du2024A,Li2024,Yan J2022,Teng 2024,Almheiri A 2020,Bueley K 2022,Wu S M 2022,Li S H 2024,Wu S M2022,Haddadi S2024}. Wang \emph{et al.} \cite{Wang2016} investigated the dynamics of quantum coherence under Unruh thermal noise and shown that the robustness of basis-dependent coherence is better than entanglement under the influence of the atom-field interaction for an extremely large acceleration. A generalisation of this result to tripartite system was performed by Ref. \cite{He2018}. Huang \emph{et al.} \cite{Huang2018} explored basis-dependent coherence of fermionic system in non-inertial frame and revealed the cohering power and decohering power of Unruh channel. Furthermore, Kollas \emph{et al.} \cite{Kollas2023} investigated cohering and decohering power of massive scalar fields under instantaneous interactions. Harikrishnan \emph{et al.} \cite{Harikrishnan2022} analysed basis-dependent coherence of a multipartite system of fermionic system in non-inertial frame.

Inspired by above the effect of relativistic effects for basis-dependent coherence, we wonder how the effect of relativistic effects for basis-independent coherence. It is an interesting questions since basis-independent measures aim to capture the intrinsic coherence of a quantum system, regardless of the choice of measurement basis \cite{Designolle2021,Radhakrishnan2019,Yin2022,Du2024}. Unlike basis-dependent coherence, which may vary with the observer’s reference frame, basis-independent coherence may offer a more robust characterization of quantum properties in relativistic settings. Exploring how these measures behave under relativistic effects, such as the Unruh effect or in curved spacetime, could unveil new aspects of quantum resource theory. Therefore, we may expect to find some interesting results.

This paper contributes to the above topic. We will investigate the effect of Unruh effect for basis-independent coherence in non-inertial frame and shown that basis-independent coherence is more robust than the basis dependent coherence in non-inertial frame.

\section{Basis-independent coherence in inertial frames}\label{sec3}
We assume that Alice and Bob initially share an entangled state in an inertial frame,
\begin{equation}
|\Phi\rangle_{AB} = \frac{1}{\sqrt{2}}\left( |0\rangle_A |0\rangle_B + |1\rangle_A |1\rangle_B \right).
\label{eq:initial}
\end{equation}
After sharing their own qubit, Bob moves with respect to Alice in uniform accelerations $a$. Using the single-mode approximation, Bob's vacuum and one-particle states $|0\rangle$ and $|1\rangle$ in Minkowski space are transformed into \cite{Alsing2006}
\begin{align}\label{Unruhmodel}
&|0_B\rangle\to\cos(r)|0_{B_I}\rangle|0_{B_{II}}\rangle+\sin(r)|1_{B_I}\rangle|1_{B_{II}}\rangle,\\\notag
&|1_B\rangle\to|1_{B_I}\rangle|0_{B_{II}}\rangle,
\end{align}
where $r$ are the acceleration parameters with the range $0\leqslant r\leqslant \pi / 4$ for  $0\leqslant a\leqslant \infty$, and  $|n_{\mathrm{B_I}}\rangle$ and  $| n_{\mathrm{B_{II}}}\rangle$$(n=0,1)$ are the mode decomposition of $|n_\mathrm{B}\rangle$ in the two causally disconnected regions I and II in Rindler space. This implies that the vacuum state defined by inertial observers for a localized system is inequivalent to the vacuum state perceived by uniformly accelerated observers (Rindler observers), as discussed in Refs. \cite{Alsing2012,Bruschi2012,Downes2011}. Using Eqs. (\ref{Unruhmodel}), we obtain
\begin{equation}
\begin{aligned}
|\Phi\rangle_{A,B_I,B_{II}} &= \frac{1}{\sqrt{2}} \cos r\, |0\rangle_A |0\rangle_{B_I} |0\rangle_{B_{II}}\\
&+ \frac{1}{\sqrt{2}}\sin r\, |0\rangle_A |1\rangle_{B_I} |1\rangle_{B_{II}}\\
&+\frac{1}{\sqrt{2}}|1\rangle_A |1\rangle_{B_I} |0\rangle_{B_{II}}.
\end{aligned}
\end{equation}
Since Bob is causally disconnected from the region $II$, the only information which is physically accessible to the observers is encoded in the mode $A$ described by Alice and the mode $I$ described by Bob. Taking the trace over the state of region $II$, we obtain
\begin{align}\label{b1}
\rho_{AB_I}=\left(
        \begin{array}{cccc}
        {\frac{1}{2}\cos^2 r} & {0} & {0} & {\frac{1}{2} \cos r} \\
        {0 }&{\frac{1}{2}\sin^2 r} & {0} & {0} \\
        {0} & {0} & {0} & {0} \\
        {\frac{1}{2} \cos r} & {0} & {0} & {1/2} \\
        \end{array}
        \right),
\end{align}
Substituting Eq. (\ref{b1}) into Eq. (\ref{coherence}), we obtain
\begin{align}
C(\rho_{AB_{I}}) =\sqrt{ S \left( \frac{\rho_{AB_{I}}+\rho_M}{2} \right) - \frac{S(\rho_{AB_{I}}) +2}{2}}
\end{align}
where
\begin{align}
S\left( \frac{\rho_{AB_{I}} + \rho_M}{2} \right)&=
\frac{3}{4}
 - \left(\frac{4 + \cos 2r}{8}\right) \log_2\left[\frac{4 + \cos 2r}{8} \right] \nonumber \\
& - \left(\frac{1 + 2 \sin^2 r}{8}\right) \log_2\left[\frac{1 + 2 \sin^2 r}{8}\right]
\end{align}
and
\begin{align}
S(\rho_{AB_I})&=-\frac{3 + \cos 2r}{4} \log_2\left(\frac{3 + \cos 2r}{4}\right)\\\notag
 &-\frac{\sin^2 r}{2}\log_2\left(\frac{\sin^2 r}{2}\right)
\end{align}
\begin{figure}
    \centering
    \includegraphics[width=0.5\textwidth]{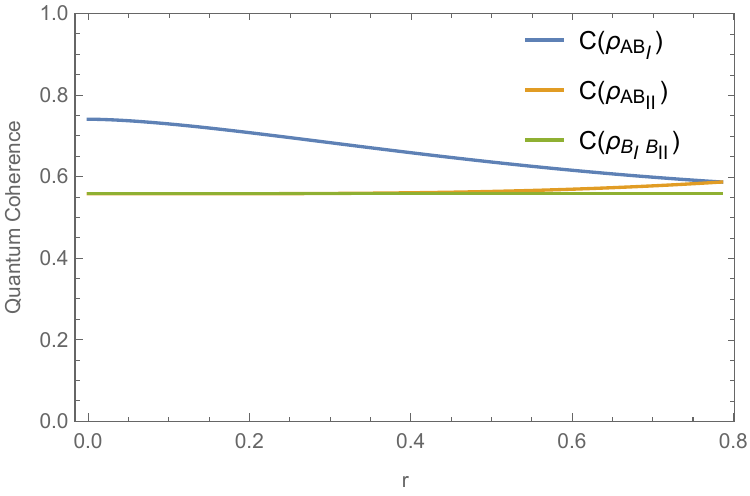}
    \caption{ Basis-independent coherence as a function of $r$.}\label{fig1}
\end{figure}
In Fig. (\ref{fig1}), we plot $C(\rho_{AB_{I}})$ as a function of $r$. For vanishing acceleration $r=0$, $C(\rho_{AB_{I}})\approx0.7407$. As the acceleration increases, the basis-independent coherence $C(\rho_{AB_{I}})$ monotonous decrease. It means that the intrinsic coherence of state $\rho_{AB_{I}}$ decreases due to the thermal fields generated by the Unruh effect. In the acceleration limits $r=\pi/4$, we have $C(\rho_{AB_{I}})\approx0.5869$, implying that the basis-independent coherence in the infinite acceleration limit is finite. This means that the state have always intrinsic coherence and can be used as a resource for performing certain quantum information processing tasks.

To explore basis-independent coherence in this system in more detail we consider the tripartite system consisting of the modes $A$, $I$, and $II$. We therefore calculate the basis-independent coherence in all possible bipartite divisions of the system.

After taking the trace over modes $B_{I}$, we obtain the density operator $\rho_{AB_{II}}$ as
\begin{align}\label{b2}
\rho_{AB_{II}}=\left(
        \begin{array}{cccc}
        {\frac{1}{2}\cos^2 r}  & {0} & {0} & {0} \\
        {0 }& {\frac{1}{2}\sin^2 r} & {\frac{1}{2} \sin r} & {0} \\
        {0} & {\frac{1}{2} \sin r} & {1/2} & {0} \\
        {0} & {0} & {0} & {0} \\
        \end{array}
        \right).
\end{align}
At zero acceleration, we obtain that
\begin{align}\label{b2}
\rho_{AB_{II}}|_{r=0}=\left(
        \begin{array}{cccc}
        {\frac{1}{2}}  & {0} & {0} & {0} \\
        {0 }& 0 & 0 & {0} \\
        {0} & 0 & {\frac{1}{2}} & {0} \\
        {0} & {0} & {0} & {0} \\
        \end{array}
        \right).
\end{align}
Since the off-diagonal elements of the density matrix are zero, the basis-dependent coherence is zero\cite{Baumgratz2014}. By substituting Eq. (\ref{b2}) into Eq. (\ref{coherence}), we obtain
\begin{align}
C(\rho_{AB_{II}}) = \sqrt{ S \left( \frac{\rho_{AB_{II}}+\rho_M}{2} \right) - \frac{S(\rho_{AB_{II}}) +2}{2}}
\end{align}
where
\begin{align}
S\left( \frac{\rho_{AB_{II}} + \rho_M}{2} \right) &=
\frac{3}{4}
- \frac{1 + 2\cos^2 r}{8} \log_2\left( \frac{1 + 2\cos^2 r}{8} \right)\\\notag
&- \frac{4 - \cos2r}{8} \log_2\left( \frac{4 - \cos2r}{8} \right)
\end{align}
and
\begin{align}
S(\rho_{AB_{II}})&=-\frac{\cos^2 r}{2} \log_2\left(\frac{\cos^2 r}{2}\right) \\\notag
& - \frac{3 - \cos 2r}{4} \log_2\left(\frac{3 - \cos 2r}{4}\right).
\end{align}
Interestingly, we find that for vanishing acceleration $r=0$, $C(\rho_{AB_{II}})=C(\rho_{AB_{I}})\approx0.5579$. This is different from the behavior of basis-dependent coherence. $C(\rho_{AB_{II}})$ increases as the acceleration increases[Showed in Fig. (\ref{fig1})]. In the infinite-acceleration limit, $C(\rho_{AB_{II}})=C(\rho_{AB_{I}})\approx0.5869$.

Tracing over the modes in $A$, we obtain the density matrix
\begin{align}\label{b3}
\rho_{B_IB_{II}}=\left(
        \begin{array}{cccc}
        {\frac{1}{2}\cos^2 r}  & {0} & {0} & {\frac{\sin r \cos r}{2}} \\
        {0 }& {0} & {0} & {0} \\
        {0} & {0} & {1/2} & {0} \\
        {\frac{\sin r \cos r}{2}} & {0} & {0} & {\frac{1}{2}\sin^2 r} \\
        \end{array}
        \right)
\end{align}

Substituting Eq. (\ref{b3}) into Eq. (\ref{coherence}), we obtain
\begin{align}\label{cb3}
C(\rho_{B_IB_{II}}) =\sqrt{-\frac{3}{4}\left(1+\log_2\left(\frac{3}{8}\right)\right)}.
\end{align}

From Eq. (\ref{cb3}), we can see that $C(\rho_{AB_{II}})$ is independent $r$ and hence the basis independent coherence is frozen[Showed in Fig. (\ref{fig1})].

\section{Conclusion}\label{sec5}
In this work, we have explored the behavior of basis-independent quantum coherence in a relativistic setting by considering a bipartite entangled Dirac field shared between an inertial observer (Alice) and a uniformly accelerated observer (Bob). By incorporating the Unruh effect into our analysis, we examined how basis-independent coherence evolves with acceleration, in contrast to the well-studied basis-dependent coherence. It is shown that: (i) The basis-independent coherence between the accessible modes \(A\) and \(B_I\) decreases with increasing acceleration; (ii) The coherence between mode \(A\) and the inaccessible region \(B_{II}\) is nonzero even at zero acceleration and increases with acceleration. This behavior is markedly different from that of basis-dependent coherence; and (iii) The coherence between the Rindler modes \(B_I\) and \(B_{II}\) remains completely unaffected by acceleration, demonstrating a "freezing phenomenon".

Future work may extend this framework to curved spacetime backgrounds \cite{Du2024A,Haddadi S2024} and multipartite systems \cite{He2018,Harikrishnan2022}, which could further enrich our understanding of coherence and quantum information in relativistic and gravitational settings.

\begin{acknowledgments}
This work was supported by the National Natural Science Foundation of China(Grant Nos. 12175106 and 92365110), the Natural Science Foundation of Jiangsu Province, China(Grant No.BK20240612), the Natural Science Research Start-up Foundation of Recruiting Talents of Nanjing University of Posts and Telecommunications(Grant No. NY222123), and the Natural Science Foundation of Nanjing University of Posts and Telecommunications(Grant No. NY223069).
\end{acknowledgments}

\end{document}